\documentclass[useAMS,usenatbib,usegraphicx]{mn2e}
\usepackage{amsmath}
\usepackage{amsfonts}
\usepackage{amssymb}

\title[Low frequency radio spectrum and spectral turnover of LS 5039]
{Low frequency radio spectrum and spectral turnover of LS 5039}

\author[Sagar Godambe  Subir Bhattacharyya  Nilay Bhatt  Manojendu Choudhury]
{Sagar Godambe$^{1}$\thanks{E-mail:sagar@physics.utah.edu}   
 Subir Bhattacharyya$^{2}$ \thanks{E-mail:subirb@barc.gov.in }
 Nilay Bhatt $^{2}$  Manojendu Choudhury$^{3}$
 \\
$^{1}$Department of Physics, Utah University, Salt Lake City, Utah 84112-0830, USA \\
$^{2}$Astrophysical Sciences Division, Bhabha Atomic Research Centre, Mumbai 400085, India \\
$^{3}$Axiom Education Pvt. Ltd., 512 Midas, M.V. Road, J.B. Nagar, Andheri(E), Mumbai 400059, India
}

\begin{document}


\pagerange{\pageref{firstpage}--\pageref{lastpage}} \pubyear{2008}

\maketitle

\label{firstpage}

\begin{abstract}
LS 5039, a possible black hole x-ray binary, was recently observed with
Giant Meterwave Radio Telescope. The observed spectrum presented here
shows that the spectrum is inverted at the low frequency. When combined
with the archival data with orbital phase similar to the present observations,
it shows a clear indication of a spectral turnover. The combined data are fitted
with a broken power-law and the break frequency signifies a possible spectral
turnover of the spectrum around 964 MHz. Truly simultaneous observations in
radio wavelength covering a wide range of frequencies are required to fix the
spectrum and the spectral turn over which will play a crucial role in developing
a deeper understanding of the radio emitting jet in LS 5039.
\end{abstract}


\section{Introduction}
LS5039 is a X-ray binary system consisting of a compact object  
and a companion star of ON6.5V(f) type with mass 22.9M$_{\odot}$. 
The mass of the compact object is not accurately known. But recent 
study shows that the mass of the compact object is 3.7M$_{\odot}$ (\cite{cas05}).
Although there are lot of debates regading the true nature of LS 5039 (\cite{mira06}),
the dynamical mass estimate indicates that it is a possible black hole x-ray binary candidate.
The orbital period of the binary system is 3.9 days (\cite{cas05}). The binary orbit
is highly eccentric with an eccentricity of 0.35 and has an inclination
angle 24$^0$.9$\pm$2$^0$.8 (\cite{cas05}).

LS5039 was proposed to be a very high energy gamma-ray emitter by \cite{parde00} and 
was recently seen at TeV $\gamma$-ray energies by gamma-ray telescope
High Energy Stereoscopic System (\cite{aha06}). The gamma-ray light curve shows 
an orbital modulation of $\sim3.9$ days. The gamma-ray flux as well as the spectral
hardness vary with the orbital phase. The flux variation with orbital phase
can be accounted for by 
considering the absorption of gamma-rays due to $\gamma\gamma$ pair production.
This also implies that $\gamma$-photons are emitted from a region within 1 AU
of the compact object (\cite{aha06}). The observed spectrum is very steep at
the superior conjunction and becomes harder as the compact object moves away
from it and the spectral index goes below 2.0 as the compact object reaches the
inferior conjuntion. But this variation of spectral hardness can not be adequately
explained by the gamma-ray absorption via pair production alone.  

This source was observed in X-ray by $ROSAT$ (\cite{mot97}), $RXTE$ (\cite{ribo99}), 
$ASCA$ (\cite{mart05}), $BeppoSAX$ (\cite{reig03}), $Chandra$ (\cite{bosc05}) and 
$XMM-Newton$ (\cite{mart05}) missions during 1996 to 2003. These observations were effectively
carried out during different orbital phases of the binary system. All these 
observations revealed different flux levels and different spectral indices
in the X-ray energy band, but no x-ray spectral state transition were reported 
in these observations. More recently \cite{bosc05} reported
results of $RXTE$ observations when the source was observed consecutively for
four days in July 2003. The X-ray flux is found to vary with the orbital phases
and it maximizes near periastron passage. This phase dependence of X-ray flux is
due to the motion of the compact object accreting matter from the wind of the
massive companion while moving in a highly eccentric orbit (\cite{bosc05}). But the observed 
anti-correlation between the photon index and the X-ray flux does not fit to
the scenario where X-rays are considered to be produced in and around an accretion disc.
\cite{bosc05} argued that the x-ray emission might be due to inverse Compton/synchrotron
process in a relativistic jet which might possibly explain the observed anti-correlation between the 
photon index and the observed flux. But the photon indices reported in these observations
are very similar to that generally observed in case of hard state spectrum of black hole x-ray binaries.

LS5039 was first observed by \cite{marti98} in radio with Very Large Array (VLA)
and the observation resulted a power-law spectrum with a negative power-law index.
This indicates an optically thin synchrotron emission by non-thermal electrons. This
observation also revealed a moderate variability in the radio flux. Later,  
\cite{ribo99} carried out an observation campaign in radio with VLA at 2.0, 3.6,
6.0 and 20.0 cm wavelengths and with Green Bank Interferometer (GBI) at 3.6 and 13.3 cm 
wavelengths. The observed spectrum was a power-law with a power-law index of -0.46$\pm$0.01
supporting the previous observation of \cite{marti98}. \cite{pare02} first 
resolved the source by observing with European VLBI Network (EVN) and the Multi-Element
Radio-Linked Interferometer Network (MERLIN) and it was found that LS 5039 consists of 
an asymmetric two-sided jet with an maximum extension of one side $\sim$ 1000 AU.  
Recently \cite{ribo08} studied the radio morphology
of LS5039 using Very Long Baseline Array (VLBA) and it was found that the radio
morphology changed into asymmetric from a very symmetric configuration within
five days, but the radio emission from the core did not vary appreciably.  
{\it But what is most interesting found in these radio observations is 
the optically thin non-thermal radio spectrum of the source which is in direct contradiction 
to the generally observed trend of flat or inverted radio spectrum for persistent
black hole x-ray binaries in the hard state~(\cite{fen01,gal03}). Therefore it is important to study LS5039
in the low-frequency radio band.}

We observed LS5039 at wavelengths 21 cm, 50 cm and 128 cm (1280 MHz, 614 MHz and 234 MHz respectively) 
using Giant Meterwave Radio Telescope~(GMRT). In this letter we discuss
the results of the observation. In Section 2 we discuss the observation and data reduction, results
are discussed in Section 3 and we conclude the paper in Section 4. 

\section{Observations and Data reduction}

The GMRT consist of 30 steerable antennas of 45m diameter in an approximate 
`Y' shape similar to the VLA but with each antenna in a fixed position. Fourteen 
antennas are randomly placed within a central 1 km $\times$ 1 km square (the 
`Central Square') and the remaining antennas form the irregular Y shape (six on
each arm) over a total extent of about 25 km. We refer the reader for details
about the GMRT array to http://gmrt.ncra.tifr.res.in and \cite{swarup91}. 
We observed LS 5039 at 1280 MHz on February 23, 2006 and simultaneously observed 
at 234 and 614 MHz on March 13, 2006. 

The observations were made in standard fashion, with each source observation 
(30 min) interspersed with observations of the phase calibrator (4 min) . 
The primary flux density calibrator was either 3C 48 and 
3C 286, with flux densities being on the scale of \cite{bars77}. Either of
these flux calibrator was observed for 15 min at the beginning and end of each observing
session. 

The data recorded 
with the GMRT was converted to FITS format and analyzed with the Astronomical Image 
Processing System (AIPS) using standard procedures. During the data reduction, 
we had to discard a fraction of the data affected by radio frequency interference 
and system malfunctions. After editing the data were calibrated and collapsed into
a fewer channels.
A self-calibration on the data was used to correct
for the phase related errors and improve the image quality. 

\begin{table}
\caption{Log of GMRT observations and results}
\label{Table:ObsLog}
\begin{center}
\begin{tabular}{|c|c|c|c|}
\hline
Frequency    & Flux Density	& MJD       & Phase \\
 (MHz)       & (mJy) 		&           &range  \\
\hline
 234     & 17.0$\pm$1.1     & 53788.78  &0.61 -- 0.67\\
 614     & 34.2$\pm$2.8     & 53788.78  & 0.61 -- 0.67\\
 1280    & 38.1$\pm$2.3    & 53806.73   &0.02 -- 0.06\\
\hline
\end{tabular}
\end{center}
\end{table}

\section{Results and Discussion}
LS5039 was observed simultaneously at 234 and 614 MHz during the orbital phase 
0.61--0.67 whereas the observation at 1280 MHz was carried out during the orbital 
phase 0.02--0.06. The details of the observation are tabulated in Table 1. 
As previously reported observations (\cite{marti98, ribo99, pare02}) show that the radio 
emission is core dominated, therefore the major contribution to the radio flux detected
in present observations where the source could not be resolved, is mainly due to the 
emission from the core. Given that the source did not undergo any X-ray state transition 
over the period of observation concerned, it can be safely assumed that the
variability of the source spanning this frequency range is not significant.
The plot of the spectrum is shown in Figure 1.
It is evident that the spectrum is inverted and 
it shows the trend of a possible turn over around 1000 MHz. To have a much clearer
picture of the spectrum we plot the present data with the archival data. We have chosen 
the data reported by \cite{marti98} which was obtained during the 
orbital phase 0.63 similar to that of our second spell of observation. The combined spectrum
is shown in Figure 2. The combined spectrum is fitted with a broken power-law given by
\begin{align}\label{eq:bpl}
F(\nu) =&\, F_{0}\,\nu^{-\alpha_{1}}  & \textrm{for} \; \nu < \nu_{b} \nonumber \\
	   =&\, F_{0}\,\nu_{b}^{-\alpha_{1}+\alpha_{2}} \, \nu^{-\alpha_{2}} & \textrm{for} \; \nu > \nu_{b} \nonumber
\end{align}
\begin{figure}
\begin{center}
\includegraphics[width=0.4\textwidth,height=0.35\textheight,angle=-90]{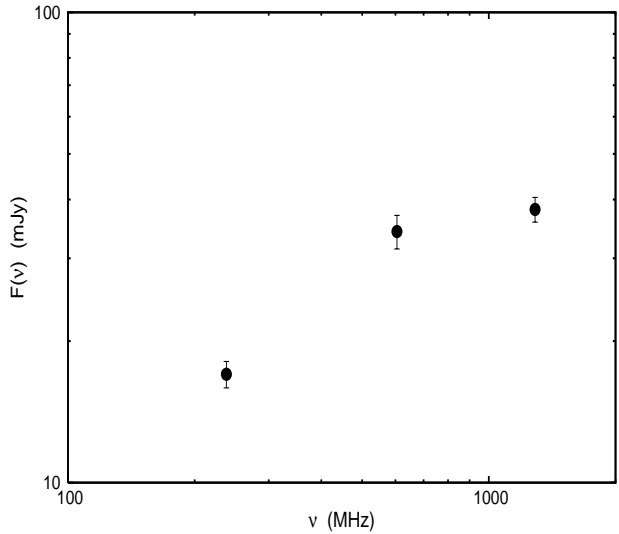}
\caption{Radio spectrum of LS 5039 as observed with GMRT}
\end{center}
\end{figure}
where, $\nu_{b}$ is the break frequency, $\alpha_{1}$ and $\alpha_{2}$ are
the power-law indices and $F_{0}$ is the normalization. We have used Levenberg-Marquardt
algorithm~(\cite{press92}) to fit the spectrum. The fitted parameters
are $F_{0}=0.282 \pm 0.183$, $\alpha_{1}=-0.749 \pm  0.111$, $\alpha_{2}=0.429 \pm  0.008$ 
and $\nu_{b}=964 \pm 104$ MHz. 
This shows that the source has an inverted spectrum with a positive spectral
index of 0.749; inverted spectrum extends upto the break frequency at 964 MHz
and then the spectrum becomes a power-law with a spectral index $-0.429$.
The power-law spectrum for frequencies greater than 964 MHz indicates optically
thin synchrotron spectrum by non-thermal electrons with spectral index 1.86. 
The break frequency at 964 MHz gives an indication that the spectrum has a
possible turnover around that frequency. This needs to be confirmed by 
further truly simultaneous observations covering the complete frequency range. 

The inverted spectrum indicates an optically thick self-absorbed synchrotron spectrum.
In ideal condition the self-absorbed, optically thick synchrotron spectrum produced
by relativistic electrons in a magnetized plasma is given by a power-law with 
index $5/2$. But in case of a magnetized plasma in a relativistic jet the physical 
condition of the plasma varies from point to point and it is very much dependent 
on the flow pattern and the shape of the jet. Therefore the resultant synchrotron 
spectrum from an unresolved jet differs substantially from the ideal situation and
the power-law index of the self-absorbed, optically thick synchrotron spectrum differs
from $5/2$. While the model by \cite{bk79} or disk--jet symbiosis
model by \cite{falbier95} explain the flat spectrum radio sources, \cite{falc96}
discussed an extended disk-jet symbiosis model to explain the inverted spectra 
from Sgr $A^{*}$ and M81$^{*}$. They extended the disk-jet symbiosis model 
by including a longitudinal pressure gradient which induces a moderate acceleration to 
a free conical jet along its flow direction.This longitudinal expansion leads to energy
losses due to adiabatic expansion and makes the proper velocity of the jet dependent
on its distance from the central engine. This model explains the inverted spectra
of Sgr $A^{*}$ (\cite{falc96}) and M81$^{*}$ (\cite{falmar00}) and may also be applicable to 
radio cores in general (\cite{falc96}). Moreover, it predicts, for a maximal jet (where 
the internal power of the jet equals the kinetic power), the slope of the inverted spectra
for different inclination angle of the source. As the microquasars are 
scaled down versions of the radio galaxies, the generic model by \cite{falc96} can be
used to model the inverted spectrum of LS5039 presented here. But 
a detailed theoretical modelling which is required to constrain the 
jet parameters and the radiative processes active in the source, is beyond the scope
of the present paper.   

\begin{figure}
\begin{center}
\includegraphics[width=0.4\textwidth,height=0.35\textheight,angle=-90]{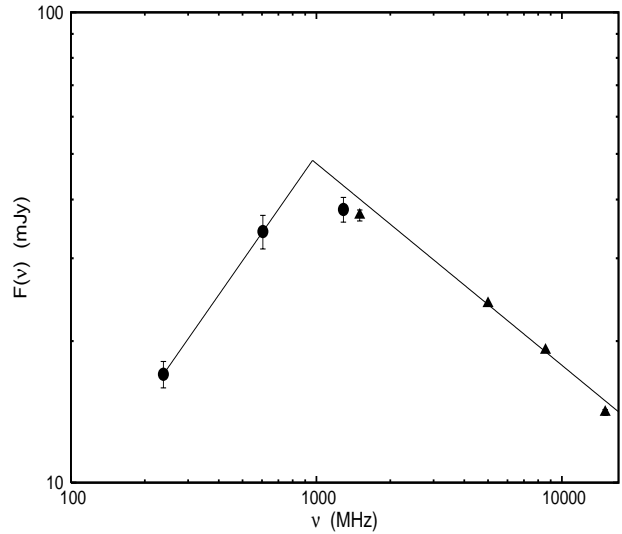}
\caption{Broken power-law fitting of the combined data set
({\it filled circle} : present data, {\it filled triangle} : 
archival data from Mart{\'{\i}} et al.~(1998))}
\end{center}
\end{figure}
\section{Conclusion}
Here, we report the low frequency radio spectrum of a X-ray binary LS5039 as observed
with GMRT. The spectrum is inverted and shows
an indication of spectral turnover. With the assumption that the variability of 
radio emission from the source is not appreciable, present data are plotted with
the archival data. The fitting of the data with a broken power-law reveals a possible indication of a spectral 
turnover at 964 MHz and an inverted spectrum with spectral index 0.749. 
This estimation of turnover frequency is indeed obtained using the archival VLA data (\cite{marti98}) with
the assumption that the source is not variable. But this estimation may depend on the variability of the
source. To have an idea about the sensitivity of the spectral break frequency on flux variability,
we introduced a moderate ($\sim 10\%$) flux variation to VLA data and fitted the spectrum. The spectral 
fitting shows that the break frequency $\nu_b$ does not change appreciably and it is well within the
uncertainities of the statistical fit.

As LS 5039 is a possible black hole x-ray binary persistently in low/hard x-ray spectral state,
the results of the present observation are consistent with the trend of inverted radio
spectrum observed for persistent black hole x-ray binary and this is the first source
of this kind for which the indication of a spectral turnover is obtained.
The spectral turnover can be used
to determine the magnetic field in the jet provided the size of the source is known
at that frequency. But it is to be noted that the magnetic field depends on the
fourth power of the angular size. Therefore, a reliable measurement of magnetic 
field is possible if and only if the angular size of the source at the turnover frequency
is determined with sufficient accuracy.
This result will further help to constrain the theoretical models to
explain the broadband spectrum from the source.

To have a deeper understanding on the radio spectrum and its dependence on the orbital
phases it is important to have simultaneous observations over the entire radio band.
Nevertheless, we are conducting a long
term campaign to study the spectral behaviour at frequencies 234, 614 and 1280 MHz over
different orbital phases of LS5039. This will allow us to constrain the low
frequency end of the spectrum of LS5039.

\section{Acknowledgement}
We acknowledge the referee, J. Mart{\'{\i}} for his comments to improve the content
of the paper.

\noindent The GMRT is a national facility operated by the National Centre for Radio Astrophysics of 
the Tata Institute of Fundamental Research. We acknowledge the help provided by the supporting 
staff of GMRT during the observations. SG also thanks C.H. Ishwara Chandra,
Chiranjib Konar and Nirupam Roy for useful discussions on radio data analysis.  SB and
NB acknowledge Abhas Mitra and Ramesh Koul for constant support and encouragement.

\bibliographystyle{mn2e}
\bibliography{ms-final}

\begin{thebibliography}{}

\bibitem[\protect\citeauthoryear{{Aharonian}, {et al.} \& {MAGIC
  collaberation}}{{Aharonian} et~al.}{2006}]{aha06}
{Aharonian} F.,  {et al.}   {MAGIC collaberation} 2006, A\&A, 460, 743

\bibitem[\protect\citeauthoryear{{Baars}, {Genzel}, {Pauliny-Toth} \&
  {Witzel}}{{Baars} et~al.}{1977}]{bars77}
{Baars} J.~W.~M.,  {Genzel} R.,  {Pauliny-Toth} I.~I.~K.,    {Witzel} A.,
  1977, A\&A, 61, 99

\bibitem[\protect\citeauthoryear{{Blandford} \& {Konigl}}{{Blandford} \&
  {Konigl}}{1979}]{bk79}
{Blandford} R.~D.,  {Konigl} A.,  1979, ApJ, 232, 34

\bibitem[\protect\citeauthoryear{{Bosch-Ramon}, {Paredes}, {Rib{\'o}},
  {Miller}, {Reig} \& {Mart{\'{\i}}}}{{Bosch-Ramon} et~al.}{2005}]{bosc05}
{Bosch-Ramon} V.,  {Paredes} J.~M.,  {Rib{\'o}} M.,  {Miller} J.~M.,  {Reig}
  P.,    {Mart{\'{\i}}} J.,  2005, ApJ, 628, 388

\bibitem[\protect\citeauthoryear{{Casares}, {Rib{\'o}}, {Ribas}, {Paredes},
  {Mart{\'{\i}}} \& {Herrero}}{{Casares} et~al.}{2005}]{cas05}
{Casares} J.,  {Rib{\'o}} M.,  {Ribas} I.,  {Paredes} J.~M.,  {Mart{\'{\i}}}
  J.,    {Herrero} A.,  2005, MNRAS, 364, 899

\bibitem[\protect\citeauthoryear{{Falcke}}{{Falcke}}{1996}]{falc96}
{Falcke} H.,  1996, ApJ, 464, L67+

\bibitem[\protect\citeauthoryear{{Falcke} \& {Biermann}}{{Falcke} \&
  {Biermann}}{1995}]{falbier95}
{Falcke} H.,  {Biermann} P.~L.,  1995, A\&A, 293, 665

\bibitem[\protect\citeauthoryear{{Falcke} \& {Markoff}}{{Falcke} \&
  {Markoff}}{2000}]{falmar00}
{Falcke} H.,  {Markoff} S.,  2000, A\&A, 362, 113

\bibitem[\protect\citeauthoryear{{Fender}}{{Fender}}{2001}]{fen01}
{Fender} R.~P.,  2001, MNRAS, 322, 31

\bibitem[\protect\citeauthoryear{{Gallo}, {Fender} \& {Pooley}}{{Gallo}
  et~al.}{2003}]{gal03}
{Gallo} E.,  {Fender} R.~P.,    {Pooley} G.~G.,  2003, MNRAS, 344, 60

\bibitem[\protect\citeauthoryear{{Mart{\'{\i}}}, {Paredes} \&
  {Rib{\'o}}}{{Mart{\'{\i}}} et~al.}{1998}]{marti98}
{Mart{\'{\i}}} J.,  {Paredes} J.~M.,    {Rib{\'o}} M.,  1998, A\&A, 338, L71

\bibitem[\protect\citeauthoryear{{Martocchia}, {Motch} \&
  {Negueruela}}{{Martocchia} et~al.}{2005}]{mart05}
{Martocchia} A.,  {Motch} C.,    {Negueruela} I.,  2005, A\&A, 430, 245

\bibitem[\protect\citeauthoryear{{Mirabel}}{{Mirabel}}{2006}]{mira06}
{Mirabel} I.~F.,  2006, Science, 312, 1759

\bibitem[\protect\citeauthoryear{{Motch}, {Haberl}, {Dennerl}, {Pakull} \&
  {Janot-Pacheco}}{{Motch} et~al.}{1997}]{mot97}
{Motch} C.,  {Haberl} F.,  {Dennerl} K.,  {Pakull} M.,    {Janot-Pacheco} E.,
  1997, A\&A, 323, 853

\bibitem[\protect\citeauthoryear{{Paredes}, {Mart{\'{\i}}}, {Rib{\'o}} \&
  {Massi}}{{Paredes} et~al.}{2000}]{parde00}
{Paredes} J.~M.,  {Mart{\'{\i}}} J.,  {Rib{\'o}} M.,    {Massi} M.,  2000,
  Science, 288, 2340

\bibitem[\protect\citeauthoryear{{Paredes}, {Rib{\'o}}, {Ros}, {Mart{\'{\i}}}
  \& {Massi}}{{Paredes} et~al.}{2002}]{pare02}
{Paredes} J.~M.,  {Rib{\'o}} M.,  {Ros} E.,  {Mart{\'{\i}}} J.,    {Massi} M.,
  2002, A\&A, 393, L99

\bibitem[\protect\citeauthoryear{{Press}, {Teukolsky}, {Vetterling} \&
  {Flannery}}{{Press} et~al.}{1992}]{press92}
{Press} W.~H.,  {Teukolsky} S.~A.,  {Vetterling} W.~T.,    {Flannery} B.~P.,
  1992, {Numerical recipes in FORTRAN. The art of scientific computing}.
Cambridge: University Press, |c1992, 2nd ed.

\bibitem[\protect\citeauthoryear{{Reig}, {Rib{\'o}}, {Paredes} \&
  {Mart{\'{\i}}}}{{Reig} et~al.}{2003}]{reig03}
{Reig} P.,  {Rib{\'o}} M.,  {Paredes} J.~M.,    {Mart{\'{\i}}} J.,  2003, A\&A,
  405, 285

\bibitem[\protect\citeauthoryear{{Rib{\'o}}, {Paredes}, {Mold{\'o}n},
  {Mart{\'{\i}}} \& {Massi}}{{Rib{\'o}} et~al.}{2008}]{ribo08}
{Rib{\'o}} M.,  {Paredes} J.~M.,  {Mold{\'o}n} J.,  {Mart{\'{\i}}} J.,
  {Massi} M.,  2008, A\&A, 481, 17

\bibitem[\protect\citeauthoryear{{Rib{\'o}}, {Reig}, {Mart{\'{\i}}} \&
  {Paredes}}{{Rib{\'o}} et~al.}{1999}]{ribo99}
{Rib{\'o}} M.,  {Reig} P.,  {Mart{\'{\i}}} J.,    {Paredes} J.~M.,  1999, A\&A,
  347, 518

\bibitem[\protect\citeauthoryear{{Swarup}, {Ananthakrishnan}, {Kapahi}, {Rao},
  {Subrahmanya} \& {Kulkarni}}{{Swarup} et~al.}{1991}]{swarup91}
{Swarup} G.,  {Ananthakrishnan} S.,  {Kapahi} V.~K.,  {Rao} A.~P.,
  {Subrahmanya} C.~R.,    {Kulkarni} V.~K.,  1991, Current Science, 60, 95

\end{thebibliography}

\end{document}